\documentclass[12pt]{article}

\usepackage{latexsym,epsf,graphicx}

\begin{document}

\newcommand{\x}{{\bf x}}
\renewcommand{\k}{{\bf k}}

\begin{titlepage}
\begin{flushright}
{October 2001}
\end{flushright}
\vspace{1cm}

\begin{center}
{\large {\bf RENORMALIZATION GROUP ANALYSIS OF A CONFINED, INTERACTING
 BOSE GAS}}
\vskip 0.5cm
G. Alber and G. Metikas

Abteilung f\"ur Quantenphysik, Universit\"at Ulm, D-89069 Ulm, Germany
\end{center}

\vspace{1cm}

\abstract{The renormalization group is not only a powerful method for describing universal properties
of phase transitions  but it is also useful
for evaluating non-universal thermodynamic properties beyond
 mean-field theory.
In this contribution we concentrate
on
these latter aspects of the renormalization group approach. We introduce its main underlying ideas
in the familiar context of the ideal Bose gas and then
apply them to
the case of an interacting, confined Bose gas within the framework of the random phase approximation. We model 
confinement by periodic boundary conditions 
and demonstrate how confinement modifies the flow equations of
 the renormalization group changing thus the thermodynamic properties of 
the gas.}

\end{titlepage}

\section{Introduction}

The recent experimental realization of Bose-Einstein condensation has renewed
the interest in the study of trapped,
dilute, weakly interacting Bose gases. 
Despite the long history of this subject there still exist unresolved 
theoretical problems for the interacting Bose gas even without the
presence of a trap. Here we
discuss some of these problems, their origins, and possible ways of overcoming 
them.

A trivial example of Bose-Einstein condensation occurs, of course, in the
non-interacting (ideal) case. What is of real physical interest though is the 
interacting case. This is so not only
because the ideal gas is, as the name suggests, an idealized approximation to
reality but also because it is the interactions that render the condensate a
true superfluid with remarkable properties \cite{FetterWalecka}.
The presence of interactions, even weak ones, complicates the mathematical
treatment of the problem and calls for approximate schemes. The most
well-known is the Bogoliubov approach \cite{Bogoliubov} and, 
though it was proposed more 
than 50 years ago, it remains the approach used most frequently in the
literature of the subject.  

The first step of the Bogoliubov approach is to simplify the interparticle
interactions. We consider that the origin of the interactions is two-body 
collisions only. The gas is
assumed to be very dilute, which is equivalent to assuming that the
interactions are very weak. This means that the interacting gas differs only 
slightly from the ideal gas and consequently most particles in the interacting
ground state should have zero momentum, as they do in the ideal case. In this
limit of low momenta, one can show that the interactions are dominantly of s-wave type. 
Thus they can be
characterized by a scattering length $a$ which
is positive for repulsive interactions. 

The second step is to replace the creation and annihilation operators
of particles in the zero-momentum state
$a^{+}_{0}$ and $a_{0}$, respectively, by their expectation
values. The reason is that we will want to treat the interaction term 
perturbatively. However, for bosons, the usual form of perturbation theory 
fails, because neither $a_{0}^{+}$ nor  $a_{0}$ annihilate the
interacting ground state. This means that it is impossible to define 
normal-ordered products with vanishing ground-state expectation value 
and consequently Wick's theorem, a cornerstone of conventional perturbation theory,
cannot be applied in a
straightforward way \cite{FetterWalecka}.
This problem is fixed if one replaces 
$ a^{+}_{0} \rightarrow \langle a^{+}_{0} \rangle = \sqrt{N_{0}} $
and $ a_{0} \rightarrow \langle a_{0} \rangle = \sqrt{N_{0}} $,
where  $N_{0}$ is the average number of condensate atoms, and reinterprets
 $ a^{+}_{0}$, $a_{0}$ as ordinary numbers that
 commute. Approximating these operators this way is justified because 
their commutator is of order $1 / \sqrt{N_{0}}$ with respect to their 
individual matrix elements and because we assume that the condensate
 density $n_{0} = N_{0}/V $, $V$ being the volume of the system, remains
 finite as $ N_{0}, V \rightarrow \infty$ in the
thermodynamic limit. In fact, this replacement leads to the so-called,
Hartree-Fock-Bogoliubov (HFB)  theory \cite{FermiBurnett}. One can do better than this 
by replacing the operators by their average value (which is an ordinary
 number) plus a small fluctuation (which is an operator), i.e. 
 $ a^{+}_{0} \rightarrow \sqrt{N_{0}} + \delta  a^{+}_{0}$,  
 $ a_{0} \rightarrow \sqrt{N_{0}} + \delta  a_{0}$,  and minimizing the 
 Hamiltonian with respect to the number of condensate atoms $N_{0}$. 
This leads to the Bogoliubov theory proper (B) \cite{FermiBurnett}, whose results, of course,
 reduce to those of HFB when the fluctuations are set to zero 
\cite{FetterWalecka,FermiFetter}. From these considerations it is
 apparent that the Bogoliubov theory is of mean-field
type, so one could in principle improve upon it by using more sophisticated
techniques. One promising possibility, near the critical region, 
is renormalization group \cite{FisherHohenberg,RSFW,StoofBijlsmaRen,Alber}.

What makes the use of beyond mean-field theory methods imperative though is 
that the Bogoliubov theory has other shortcomings in addition to its
mean-field character \cite{Huang}.
In the critical region the Bogoliubov theory 
simply does not work because there are fluctuations around
the mean-field that cannot be treated perturbatively. This happens because,
as the temperature approaches the critical temperature $T_{c}$, the thermal
cloud density develops an infrared singularity and thus diverges as the momentum 
tends to zero
\cite{RSFW,FermiBurnett}.
This really calls for Wilsonian renormalization treatment \cite{WilsonKogut}
of the problem around the critical region. It is well-known that
renormalization techniques circumvent such infrared singularities because, to
put it simply, one arrives at the flow equations without having to integrate
out all momenta down to zero \cite{Fisher}. 

Furthermore, within the Bogoliubov theory, it is usual to
make a further approximation. One assumes that the presence of the condensate
and of the thermal cloud does not modify significantly the effective
interaction between two colliding atoms from its vacuum value. This is the 
Bogoliubov-Popov theory (BP) \cite{Popov}. 
To avoid this approximation, which is
particularly unjustified in two-dimensional cases or when the interactions
are attractive, one has to use many-body T-matrix theory 
\cite{ProukakisBurnettStoof, BijlsmaStoof}. However, this
approach also runs into infrared singularity problems in the critical region.
Therefore renormalization techniques would have to be employed to take
correctly into account the many-body effect on two-body collisions near the
critical region.  
 
However,
the renormalization group is not only a powerful method for describing interacting quantum systems
close to a
phase transitions  but it is also useful
for evaluating thermodynamic properties at arbitrary temperatures thereby transcending mean-field theory.
In this contribution we concentrate
on
these latter aspects of the renormalization group approach by introducing its main underlying ideas
in the familiar context of the ideal Bose gas and by
applying them to
the case of an interacting, confined Bose gas within the framework of the random phase approximation. 

This paper is organized as follows:  
In section 2, we present an introduction to renormalization group
methods in the familiar context of the ideal Bose gas, in an arbitrary number of spatial dimensions. 
In section 3
we apply the renormalization group to the 
realistic case of an interacting, confined Bose gas 
whose confinement can be modelled by
periodic boundary conditions. 
Comparisons with the ideal gas are made where appropriate.  
We analyse the resulting critical fixed point and its associated 
critical exponent characterizing the scaling of the 
correlation length in three spatial dimensions.
We also investigate a particular non-universal property which is accessible
to experimental observation, namely the second order
coherence factor, and discuss the effect of one-dimensional confinement 
on this physical quantity.

\section{Renormalization Group and the Ideal Bose Gas}

Before treating the interacting Bose gas case, 
we illustrate the basic ideas of 
renormalization group in the familiar context of the ideal Bose gas.\\
\\
Let us start from the path integral representation of the grand-canonical partition function of the ideal Bose gas
in $D$ dimensions
\cite{Feynman}, i.e. 
\[ Z(\mu,\beta,V) \equiv {\rm Tr} e^{-\beta(\hat{H} - \mu \hat{N})} = 
\int \delta[\phi,\phi^*] e^{-S[\phi,\phi^*]} \]
with the (dimensionless) action 
\begin{equation}
S[\phi, \phi^*] =  \frac{1}{\hbar}
 \int_0^{\hbar \beta } d\tau \int_{V} d^D {\bf x}  
\left[ \phi^* (\tau, {\bf x})
[\hbar\frac{\partial}{\partial \tau} - \frac{\hbar^2}{2m} \nabla^{2} - \mu]
 \phi(\tau, {\bf x}) \right] .
\label{idealaction}
\end{equation}
The boson mass is denoted by $m$, $\mu$ is the chemical potential and
$\beta = 1/(k_B T)$ is the inverse temperature with Boltzmann's constant $k_B$. Due to the trace operation involved in the definition of the partition 
function the complex-valued field $\phi(\tau, {\bf x})$ has to fulfill periodic boundary
conditions with respect to the imaginary time $\tau$, i.e.
$\phi(\tau, {\bf x}) = \phi(\tau + \hbar \beta, {\bf x})$.
Thermodynamic properties of the non-interacting Bose gas can be evaluated from 
the extensive quantity
\begin{equation}
F(\mu,\beta,V) = - \ln Z (\mu, \beta, V) \equiv \beta \Omega (\mu, \beta, V)= \sum_{i} \ln \left[ 1 - e^{-
    \beta ( \frac{\hbar^{2} \k_{i}^{2}}{2m} - \mu)}\right] 
\label{Func}
\end{equation}
with $V$ denoting the volume of the system. $\Omega (\mu, \beta, V)$ is
 the grand-canonical thermodynamic potential.
Modelling confinement by
periodic boundary conditions with spatial periods $L_i$ ($i \in\{1, ..., D\}$) the corresponding possible wave numbers are given by
${\bf k}_i = 2\pi n_i/L_i$ with $n_i$ being integer.

If we want to evaluate the thermodynamic function of
Eq.(\ref{Func}) numerically, it is convenient
to introduce first of all an
ultraviolet momentum cutoff $|\hbar{\bf k}_{\rm max}|
\equiv \hbar \Lambda$ which
constitutes an upper limit for 
the summation over all states of the ideal Bose gas.
In order to ensure convergence of the partition function
this momentum cutoff has to be much larger than all physical momenta
characterizing the problem, i.e. $(\hbar \Lambda)^2/(2m) \gg 1/\beta, \mu$.
Introducing this ultraviolet momentum cutoff and the dimensionless scale
parameter $l$ according to the relation
$|\k | = \Lambda e^{-l}$, in the continuum limit, i.e. 
$L_i \to \infty$, Eq.(\ref{Func}) 
can be approximated by 
\begin{equation}
F(M_0,b_0,V) \equiv \int_{0}^{\infty}~dl'~{\cal N}~e^{-Dl'}~
 \ln\left[1 - e^{ - b(l') ( \frac{1}{2} 
  - M(l'))}\right] 
\label{idealF}
\end{equation}
with the 
dimensionless, rescaled thermodynamic parameters
\begin{eqnarray}
b(l) &=& b_0 ~e^{-2l} \equiv (\beta/\beta_{\Lambda})~e^{-2l} \nonumber\\
M(l) &=& M_0 ~e^{2l} \equiv (\mu \beta_{\Lambda})~ e^{2l}
\label{bM}
\end{eqnarray}
and with the cutoff dependent
scale factor $\beta_{\Lambda} = m/(\hbar \Lambda)^2$.
The quantity
${\cal N} dl' = [V \Lambda^{D} \Omega_{D} / (2 \pi )^{D}]~dl'$ 
denotes the number of states in volume $V$ within the infinitesimal
momentum shell $\hbar \Lambda (1 - dl') \leq |\hbar {\bf k}| \leq \hbar \Lambda$. $ \Omega_{D} = 2 \pi^{D/2} / \Gamma(D/2)$ is the surface of a $D$-
dimensional hypershpere of unit radius and $\Gamma(x)$ denotes the Gamma 
function \cite{Stegun}.

The main idea underlying the renormalization group approach is to perform the integration
over all states involved
in the evaluation of the partition function or of the thermodynamic
function of Eq.(\ref{idealF}) not in one step but in 
small, successive steps. 
Integrating out some of the states in one of these small steps  yields
scaling relations from which one can determine the non-analytic behaviour
of thermodynamic functions at a phase transition.
Successive applications of this partial decimation procedure eventually yields the
partition function.

In order to derive such a scaling relation for the thermodynamic function of 
Eq.(\ref{idealF}) let us integrate out a small momentum shell corresponding to the interval $l'\in [0,l]$, for example.
For the ideal Bose gas this yields the result
\begin{eqnarray}
&& F(M_{0},b_{0},V) = 
(\int_{0}^{l}~dl' + \int_{l}^{\infty}~dl')~\{{\cal N}~e^{-Dl'}~
 \ln\left[1 - e^{- b(l') ( \frac{1}{2} - M(l'))}\right] \} 
\label{scal} \nonumber \\
&& = \int_{0}^{l}~dl'~{\cal N}~e^{-Dl'}~
 \ln\left[1 - e^{- b(l') ( \frac{1}{2} - M(l'))}\right] +
e^{-Dl}~F(M(l),b(l),V). 
\end{eqnarray}
The last equality of Eq.(\ref{scal}) has been obtained by
shifting the integration variable $l'$ in the second integral according to
$l' \to l' -l$. 
The two characteristic steps of Wilsonian renormalization are 
apparent in the derivation of this
scaling relation.
The first step, 
the {\em Kadanoff transformation},
involves a partial decimation of some states of
the Bose gas. In the second step the original ultraviolet momentum 
cutoff $\hbar \Lambda$ is re-established 
by rescaling momenta according to the transformation $|\hbar {\bf k}| \to |\hbar {\bf k}|~e^{l}$.
In our case, this latter {\em trivial rescaling} is performed by the translation $l' \to l' - l$.
In configuration space
this trivial rescaling implies that we are increasing the minimum distance of resolution or effective block size
from $1/\Lambda$ to
$1/(\Lambda ~e^{-l})$ 
so that with increasing values of $l$ the physical system is described 
on larger and larger length scales.
Thus, expressed in terms of this minimum distance of resolution any physical length $L$ shrinks with increasing values
of $l$ according to $L \to Le^{-l}$.
Scaling relations of the form of Eq.(\ref{scal})
are useful for investigating the non-analytic scaling behaviour 
of thermodynamic quantities close to a second
order phase transition \cite{Huang}. 

Applying the scaling relation of Eq.(\ref{scal}) repeatedly 
the evaluation of
the thermodynamic function $F(M_0,b_0,V)$ can be reduced 
to the solution of a system of ordinary differential equations, i.e.
\begin{eqnarray}
\frac{dF(l)}{dl} &=& {\cal N} e^{-Dl} \ln\left[1 - e^{ - b(l)
(\frac{1}{2}  - M(l))}\right] ,\nonumber\\
\frac{dM(l)}{dl} &=& 2M(l),\nonumber\\
\frac{db(l)}{dl} &=& - 2b(l),
\label{idealRGb}
\end{eqnarray}
which have to be solved with the initial conditions  $M(l=0) = M_0 \ll 1$ and $b(l=0) = b_0 \gg 1$
consistent with the choice of the ultraviolet cutoff.
Thereby,
the thermodynamic function $F(M_0,b_0,V) \equiv F(l\to\infty)$
is obtained from the solution of Eqs.(\ref{idealRGb})
with the additional initial condition $F(l=0) = 0$. 
We will be referring to Eqs.(\ref{idealRGb})
as the renormalization group (RG) equations. 
We note that the equation for $F(l)$ depends on the solutions 
$M(l)$ and $b(l)$
but $F(l)$ itself does not couple back to the equations of motion for 
these two quantities.
Thus, one can first solve the two equations of motion
for the scaled thermodynamic parameters $M(l)$ and $b(l)$ and insert the resulting solutions into
the equation of motion for $F(l)$. In the case of the
ideal Bose gas
the RG equations for $M(l)$ and $b(l)$  are simple and describe
the trivial scaling of these scaled thermodynamic parameters.
This example demonstrates the basic concepts involved in the
evaluation of thermodynamic partition functions by RG methods.

\begin{figure}
\begin{center}
\includegraphics[scale=0.4]{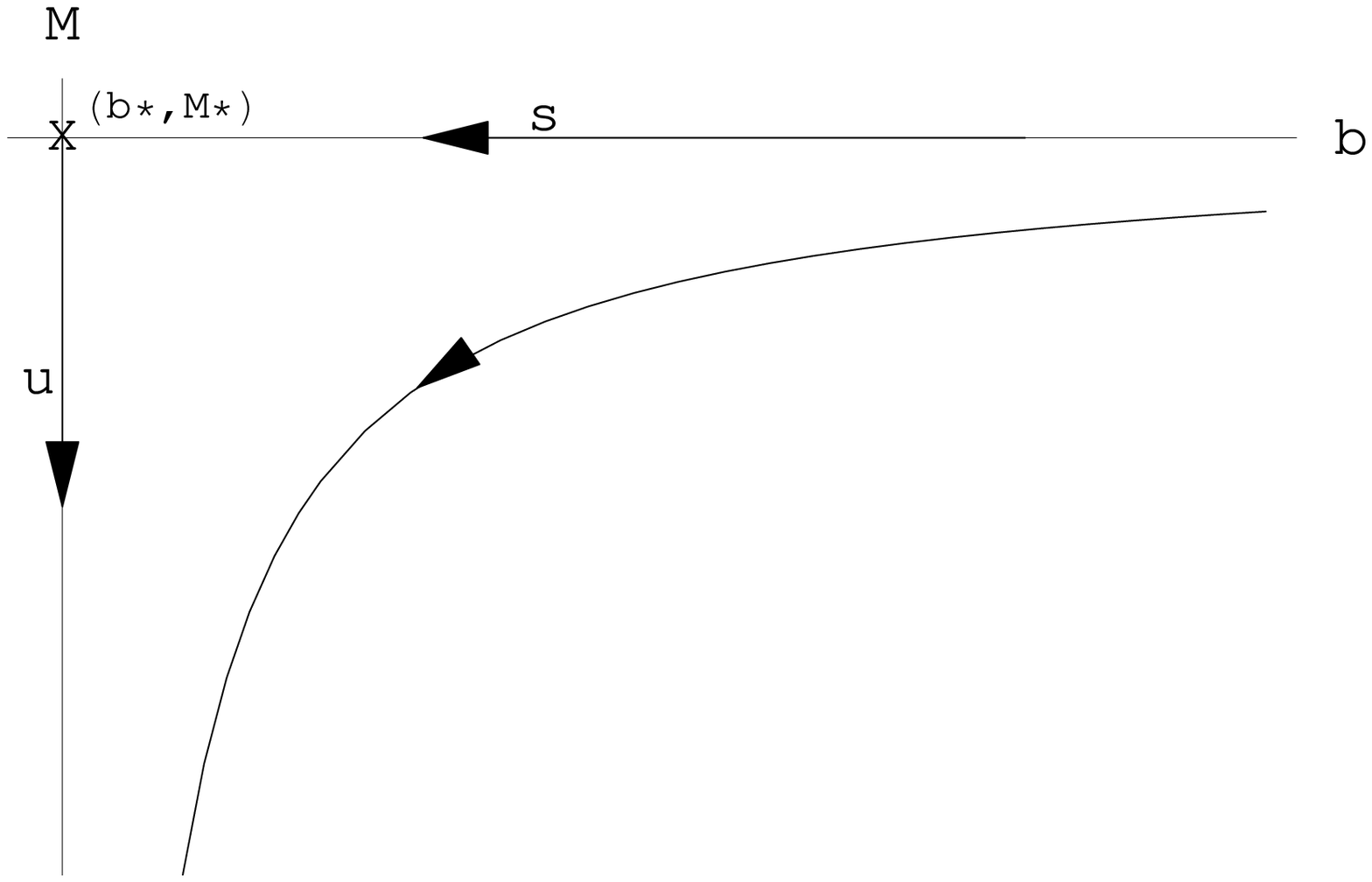}
\caption{Flow of the scaled thermodynamic parameters $(b(l),M(l))$ 
for the ideal Bose gas according to the
RG equations Eqs.(\ref{idealRGb}).
With increasing renormalization, i.e. increasing values of $l$,
an initial point $(b_0,M_0)$ with $b_0 = b(l=0)\gg 1$ and
 $M_0 = M(l=0) < 0$ in the thermodynamic parameter space  is driven to higher effective temperatures, i.e.
to smaller values of $b(l)$, and to more negative effective chemical potentials.
The unstable fixed point $(b_*,M_*) = (0,0)$ is indicated by a cross.
\label{Fig1}}
\end{center}
\end{figure}

The system of Eqs.(\ref{idealRGb}) 
is autonomous with an unstable, fixed point at $(b_{*},M_{*}) \equiv (0,0)$.
The parametric plot of a typical trajectory $M(l)$ versus $b(l)$
is drawn in Fig.\ref{Fig1} together with
this fixed point and its associated stable (s) and unstable (u) manifolds.
It is well known that in the grand-canonical ensemble the phase transition
 of the ideal Bose gas
occurs at zero chemical potential, i.e. at $M_0 = 0$. According 
to Eqs.(\ref{idealRGb})
this reflects the fact that at the critical point thermodynamic properties
 of an ideal Bose gas
are determined by the stable manifold of the unstable fixed
 point $(b_{*},M_{*})$.
The unstable manifold of this fixed point governs the behaviour
of the ideal Bose gas close to criticality.
The eigenvalues of the scaled thermodynamic parameters corresponding to the stable and unstable manifolds are given by
$\lambda_{-}=-2$ and $\lambda_{+}=2$ (compare with Eqs.(\ref{idealRGb})).
In particular, the
positive eigenvalue $\lambda_{+}=2$ which is associated with the (one dimensional)
unstable manifold 
determines the critical exponents governing
the scaling relations of singular parts of physical quantities near the critical point 
\cite{Huang,Fisher}. As an example, let us consider the correlation length $\xi_0$ of the ideal Bose gas
which changes under renormalization
according to $\xi(l) = \xi_0 ~e^{-l}$.
Furthermore, according to Eqs.(\ref{idealRGb}),
 the relevant variable scales as $(M(l) - M_*) = (M_0 - M_*)e^{\lambda_+ l}$.
If the system is close to the critical point, i.e. $\mid M_0 - M_*\mid \ll 1$, and if we iterate
the renormalization group equations up to a point $l_0 \gg 1$ far away from this critical point, say
with $M(l_0) - M_* = 1$, we find  the scaling relation
\begin{equation}
\xi_0 = \xi (l_0) e^{l_0} = \xi(l_0)(M_0 - M_*)^{-1/\lambda_+}\equiv \xi(l_0)(M_0 - M_*)^{-\nu}.
\label{corrlength}
\end{equation}
Thus, the divergence of the correlation length at the critical point is governed by the critical exponent
$\nu = 1/\lambda_+$.
With the help of scaling relations such as Eq.(\ref{scal}),
one may also relate the characteristic eigenvalue $\lambda_+$ of the unstable manifold 
to the critical exponents of other physical quantities of interest.
Let us consider the thermodynamic function $F(M_0,b_0,V)$ as a further example.
Whereas
for any finite value $l_0$ the integral in the last line of Eq.(\ref{scal}) is a smooth function of
the intensive thermodynamic parameters $\mu$ and $\beta$, the second term of this last line
gives rise to a
non-analytical behaviour
in the neighbourhood of the critical point. If we are close to the critical point and choose $l_0$ again so large
that $(M(l_0) - M_*)=1$ and $b(l_0) \ll 1$ we find that the singular part of this thermodynamic function scales as
\begin{equation}
F_s(M_0,b_0,V) = e^{-Dl_0}~F_s(1,0,V) \equiv (M_0 - M_*)^{D/\lambda_+}~F_s(1,0,V).
\end{equation}

\section{The Interacting Bose Gas}

Treating two-body collisions between bosons in the low-momentum or s-wave
approximation the
path integral representation of the partition
function of the  homogeneous interacting Bose-gas is
given by
\begin{equation}
 Z(\mu,\beta,V,g) \equiv {\rm Tr} e^{-\beta(\hat{H} - \mu \hat{N})} = 
\int \delta[\phi,\phi^*] e^{-S[\phi,\phi^*]} 
\label{partition}
\end{equation}
with the (dimensionless) action 
\begin{equation}
S[\phi, \phi^*] =  \frac{1}{\hbar}
 \int_0^{\hbar \beta } d\tau \int_{V} d^D {\bf x}  
\left[ \phi^* (\tau, {\bf x})
[\hbar\frac{\partial}{\partial \tau} - \frac{\hbar^2}{2m} \nabla^{2} - \mu]
 \phi(\tau, {\bf x}) + \frac{1}{2} g |\phi(\tau, {\bf x})|^4
 \right]. 
\label{action}
 \end{equation}
In the low-momentum approximation the interparticle
interaction can be described by the zero-momentum component of the Fourier transform of the two-body interaction potential.
Thus, within this approximation
a repulsive, short-range potential can be characterized by a positive interaction strength $g$. 
In three spatial dimensions, for example, this interaction strength is
related to the positive scattering length $a$ of the interparticle interaction by
the familiar  relation $g = 4\pi\hbar^2 a/m$. 

In the rest of this paper we will be interested not only in the case of an
unconfined interacting Bose gas but also in cases where at least one of the
spatial degrees of freedom is confined by a potential. In the simplest approximation
such a confinement in the $3$- or $z$-direction, for example,
can be described by a periodic boundary condition 
of the form
\hspace*{4cm} \[ \phi(\tau,{\bf x}) = \phi(\tau,{\bf x} + L_z {\bf e}_z)  \]
with
$L_z$ denoting the characteristic length of confinement. Within such a description of confinement 
the translational invariant character of the problem is conserved.

Before addressing the evaluation of the partition function of Eq.(\ref{partition}) with the help of
the renormalization group let us briefly summarize its approximate evaluation within
the framework of mean-field theory.
In mean-field theory one approximates the partition function of  Eq.(\ref{partition})
by saddle point integration \cite{BenderOrszag}.
For this purpose one determines first of all the most probable configuration
$\overline{\phi}({\bf x})$ by minimizing the action of Eq.(\ref{action}). This way one
arrives at the Gross-Pitaevski equation \cite{FetterWalecka}
\begin{equation}
\left[ \hbar \frac{\partial}{\partial \tau}  - \frac{\hbar^2}{2m}
\nabla^{2} - \mu + g |\overline{\phi} ({\bf x})|^2 \right] 
\overline{\phi}({\bf x}) = 0.
\label{GP}
\end{equation}
In the homogeneous case that we are examining, the most probable static and 
space-independent configuration is given by
\begin{equation}
\overline{\phi} = \sqrt{\mu/g}.
\label{mostprob}
\end{equation}
In a second step one expands the action of Eq.(\ref{action}) 
around this most probable
configuration up to second order assuming that the fluctuations around this
most probable configuration are small. The resulting Gaussian integrations can be easily
performed, thus yielding the well-known mean-field approximation for the partition function \cite{Wiegel}.

\subsection{Renormalization group approach}

Using the renormalization group it is possible to improve on this mean-field
approximation. However, contrary to the case of the ideal Bose gas,
in the presence of interparticle interactions the first step of the
renormalization group method, namely the Kadanoff transformation, can be implemented only 
approximately. A frequently employed approximation in this context is the random phase approximation \cite{Hertz}
whose resulting renormalization group equations will be discussed in the following.

\subsubsection*{Kadanoff Transformation}

As we want to take into account the fluctuations of the complex-valued field $\phi(\tau,{\bf x})$ beyond
mean-field theory, it is convenient to separate 
this field according to
\begin{equation}
\phi(\tau,{\bf x}) \to \overline{\phi} +  \phi(\tau,{\bf x}).
\label{expand}
\end{equation}
Thus, we obtain the following symmetry broken form of the (dimensionless)
 action
\begin{eqnarray}
S[\phi,\phi^{*}]=
-\beta V [\mu n_{0} - \frac{n_{0}^2 g}{2}] \hspace{-0.5cm}
 && + \frac{1}{\hbar}
\int dx~\phi^{*}(x) 
 [ \hbar~\frac{\partial}{\partial \tau} - \frac{\hbar^{2}}{2m}~\nabla^{2}
 -\mu + 2 g n_{0}] \phi (x) \nonumber \\
&& + \frac{g n_{0}}{2\hbar}~\int dx~\left[\phi^{*}(x) \phi^*(x) + 
\phi(x) \phi(x) \right] \nonumber \\
&& + \frac{g \overline{\phi}}{\hbar}~\int dx~\left[ \phi^{*}(x) 
\phi^{*}(x) \phi(x) + \phi^{*}(x)
 \phi(x) \phi(x) \right] \nonumber \\
&& + \frac{g}{2\hbar}~\int dx~\phi^{*}(x) \phi^{*}(x) \phi(x) \phi(x) 
\label{brokenaction}
\end{eqnarray}
with the condensate density
$n_{0}=|\overline{\phi}|^{2}=\mu/g$. 
The ${\rm U}(1)$ global symmetry of the action of Eq.(\ref{action})
 has been broken
spontaneously by the introduction of the most probable configuration
 $\overline{\phi}$.
In order to
simplify the notation we have taken $x$ to represent both time 
and coordinates, so, for example, $\phi(x)\equiv \phi(\tau, {\bf x})$ and
$dx\equiv d\tau~d^{D}{\bf x}$.

\begin{figure}
\begin{center}
\includegraphics[scale=0.4]{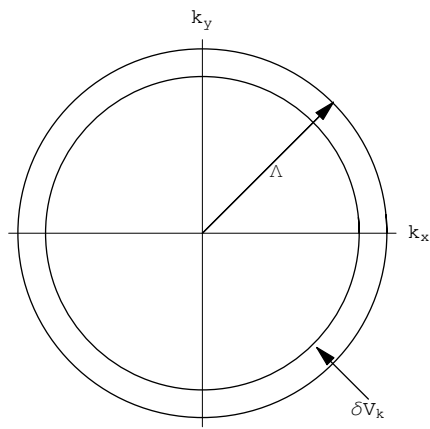}
\caption{Schematic representation of the region in momentum space which is eliminated in the Kadanoff transformation.
\label{Fig2}}
\end{center}
\end{figure}

Now we want to integrate out large momentum fluctuations, so we also
split the fluctuation $\phi(x)$ around the most probable configuration $\overline{\phi}$ into
a long wave length component $\phi_<(x)$  and into a short wave length component
$\delta \phi_>(x)$, i.e. 
\begin{equation}
\phi(x) = \phi_<(x) + \delta \phi_>(x)
\label{split}
\end{equation}
with
\begin{equation}
\phi_{<}(x)= \sum_{ \k_{{\bf m}} \in V_{k} - \delta V_{k} }
\sum_{ n \in {\bf Z} } \frac{ e^{i \k_{{\bf m}} \cdot {\bf x} } } {\sqrt{V}}
\frac{e^{-in \omega \tau } }{\sqrt{ \hbar\beta } }~\varphi_{n {\bf m}}
\end{equation}
and
\begin{equation}
\delta \phi_{>}(x)=
\sum_{ \k_{{\bf m}} \in \delta V_{k} }^{'} \sum_{ n \in {\bf Z} }
\frac{e^{i {\bf k}_{\bf m}\cdot {\bf x}}}{\sqrt{V}}
\frac{e^{-in \omega \tau } }{\sqrt{ \hbar \beta } }~\delta\varphi_{n {\bf m}},\end{equation}
where
$\varphi_{n {\bf m}}$ ($\delta \varphi_{n {\bf m}}$) are the Fourier 
components of $\phi_{<}(x)$ ($\delta \phi_{>}(x)$),
$n \omega  \equiv  n~(2 \pi/ \beta \hbar)$ are the Matsubara frequencies,
and
$\hbar V_{k}$ denotes a hypersphere of radius $\hbar \Lambda$ in
$D$-dimensional momentum space.
The short wave length fluctuations involve momentum components which are 
contained only in an infinitesimally thin shell in momentum space of 
thickness
$\hbar\Lambda(1 - dl)\leq |\hbar {\bf k}|\leq \hbar\Lambda$ and which is denoted $\hbar \delta V_{k}$ (compare with Fig.\ref{Fig2}). 
For the sake of simplicity 
we will be referring to $\phi_<(x)$ as the lower field and to
$\delta \phi_>(x)$ as the upper field.

The upper field which we want to eliminate from the partition function involves an infinitesimal momentum shell. Therefore, 
it is sufficient to expand the action of Eq.(\ref{brokenaction}) up to 
second order in terms of the upper field \cite{Wegner}. 
The resulting expression can be further simplified 
in the random phase approximation \cite{Hertz}. In this approximation it is
 assumed that the upper field is rapidly varying in space in comparison with
the lower field so that the dominant contributions to the action arise from
 those particular terms quadratic
in the upper field which are slowly varying in space. Thus, in the random phase approximation, the action reads
\begin{eqnarray}
\hspace{-1cm} S[\phi,\phi^*]  & = & S[\phi_<,\phi_<^*] + 
 \frac{1}{\hbar}\int dx~\delta \phi_{>}^{*}(x)~ 
 \left[ \hbar~\frac{\partial}{\partial \tau} - \frac{\hbar^{2}}{2m}~\nabla^{2} - \mu \right]~\delta \phi_{>}(x) \nonumber \\
&& +
\frac{2g}{\hbar}~\left[\int dx~
|\overline{\phi} + \phi_{<}(x)|^{2} \right]
~\left[\frac{1}{V \hbar \beta}\int dx~
\delta \phi_{>}^{*}(x)~ 
\delta \phi_{>}(x)\right] \nonumber \\
&& + \frac{g}{2 \hbar}~\left [\int dx~
( \overline{\phi} + \phi_{<}(x) )^{2} \right]
~\left[\frac{1}{V\hbar\beta} \int dx~\delta \phi_{>}^{*}(x) 
\delta \phi_{>}^{*}(x)\right] \nonumber \\
&& + \frac{g}{2\hbar}~\left[\int dx~(\overline{\phi} + \phi^*_<(x))^2\right]
~\left[\frac{1}{V\hbar \beta} \int dx~\delta \phi_{>}(x) 
\delta \phi_{>}(x)\right], 
\label{secondorderacrion}
\end{eqnarray}
where we have taken into account that terms linear in $\delta \phi_>(x)$ and $\delta \phi_>^*(x)$ vanish because the fluctuations are expanded around the most probable configuration $\overline{\phi}$ which fulfills the Gross-Pitaevski equation (\ref{GP}). 
Performing the Gaussian integration in terms of the upper fields yields
 an effective action for the lower field of the form \cite{Alber}
\begin{equation}
 S_{\rm eff}[ \phi_{<},\phi_{<}^*] = S[\phi_{<},\phi_{<}^*] 
 + \delta S[\phi_<,\phi_<^*].
\label{effaction}
\end{equation}
We want to point out that, contrary to the case of the mean-field approximation, now
the contribution to the effective action, i.e. $\delta S[\phi_<,\phi_<^*]$,
still
depends on the lower field which characterizes the long wave length fluctuations.
Stated differently, the mean-field approximation would correspond to the replacement $\delta S[\phi_<,\phi_<^*] 
\to \delta S[\phi_<\equiv 0,\phi_<^*\equiv 0]$ in this step of the decimation procedure.
Thus,
after integration over the short wave length fluctuations
the simplest improvement transcending the mean-field approximation is obtained by expanding
$\delta S[\phi_<,\phi_<^*]$ up to second order in the lower field. 
As apparent from Eq.(\ref{brokenaction}), such a second order expansion leads to a change of
the most probable configuration, i.e. $\overline{\phi}\to\overline{\phi} + \delta \overline{\phi}$,
through the terms linear in $\phi_<$ and $\phi_<^*$ and to a change
of the chemical potential through the terms quadratic in the lower field, i.e.
$\mu \to \delta \mu$.
Enforcing
the relation (\ref{mostprob}) at each step of the renormalization procedure implies
a corresponding scaling of the interparticle coupling strength $g$ according to
the relation
$|\overline{\phi}+\delta \overline{\phi}|^2 \equiv (\mu + \delta \mu)/(g + \delta g)$ 
\cite{StoofBijlsmaRen,Alber}.

\subsubsection*{Trivial rescaling}

The second step of the renormalization procedure is to recast 
the effective action of Eq.(\ref{effaction}) in terms of
 the new chemical potential $\mu_{eff} = \mu + \delta \mu$
and the new interparticle coupling strength $g_{eff} = g + \delta g$
 in the form of
the original action by re-establishing the original
momentum cutoff
%
%
$\hbar\Lambda$. For this purpose we have to 
rescale momenta according to $|\k | \to |\k (l) | = |\k | e^{l}$. 
As in the ideal gas case,
 this transformation and the demand that the action remains formally the same
 after each renormalization step induce the appearance of rescaled parameters 
 in the action, i.e.
\begin{eqnarray}
 V & \to & V(l)= V~e^{-Dl} \nonumber \\
 \beta & \to & \beta(l)=\beta~e^{-2l} \nonumber \\
 \phi & \to & \phi(l) = \phi~e^{-l} \nonumber \\
 \mu &\to & \mu(l)=(\mu+ \delta \mu)~e^{2l} \nonumber \\
 g &\to & g(l)= (g+\delta g)~e^{(2-D)l}.
\label{trivialscaling} 
\end{eqnarray}
These scaling relations reveal two differences from
 the corresponding scaling relations of the ideal gas.
The first is that
 there is, of course, an additional equation for the rescaling of the coupling
  constant. The second is that the rescaled chemical potential contains an
 additional correction $\delta \mu$ which itself depends on $l$ and is
 therefore a non-trivial contribution to the scaling of $\mu$. This
 non-trivial part was not present in the ideal gas case.
 We also note that the coupling constant equation contains a non-trivial
 part $\delta g$ as well.

\subsubsection*{The Confined Renormalization Group Equations}
\label{RGE}

Because we have taken the renormalization step $dl$ to be infinitesimal, the 
corrections for the chemical potential and the coupling constant take the form 
of differential equations with respect to the continuous parameter $l$. Introducing the 
dimensionless, scaled coupling constant
\begin{equation}
G(l) = \beta_{\Lambda}~\Lambda^{D}~g(l)
\end{equation}
and the scaled thermodynamic parameters of Eqs.(\ref{bM})
we obtain the renormalization group equations
for the interacting Bose gas
\cite{Alber}
\begin{eqnarray}
\frac{d F(l)}{dl} &=& V \Lambda^{D}~d(l)~e^{-Dl}~\left[ \ln \left(
2 \sinh [\lambda(l)/2 ] \right) - b(l)(\epsilon_> - M_0~e^{2l})/2  \right] ,
\nonumber\\
\frac{d M(l)}{dl} &=&  2 M(l) + d(l)~G(l)~A[M(l),b(l)],
\nonumber\\
\frac{dG(l)}{dl}  &=&  -(D-2) G(l) + d(l)~[G(l)]^2~B[M(l),b(l)],\nonumber\\
\frac{db(l)}{dl}&=&-2 b(l)
\label{RG}
\end{eqnarray}
with
\begin{eqnarray}
 A[M(l),b(l)] &=&
 b(l)~\frac{\coth[\lambda(l)/2]
}{2\lambda(l)}~[2M(l)-2\epsilon_{>}]
 - [b(l)]^{3}~\frac{M(l)}{2 \lambda(l)^2} \nonumber \\
&& \left[ \frac{1}{2 \sinh^2
(\lambda(l)/2)} + \frac{\coth[\lambda(l)/2]
}{\lambda(l)} \right]~
 [2\epsilon_{>} + M(l)]^2,\nonumber
\\
B[M(l),b(l)]&=& 3b(l)~\frac{\coth [\lambda(l)/2]
}{2 \lambda(l) }
 - [b(l)]^3~\frac{1}{2 \lambda(l)^2} \nonumber \\
&& \left[ \frac{1}{2 \sinh^{2}
[\lambda(l)/2]} + \frac{\coth [\lambda(l)/2]
}{\lambda(l)} \right] 
[2 \epsilon_{>} + M(l)]^2
\label{B}
\end{eqnarray}
and with the scaled cutoff energy $\epsilon_> = 1/2$.
The quantity
\begin{equation}  
\lambda (l)=b(l) \sqrt{\epsilon_>\left[\epsilon_> + 2M(l)\right]}
\end{equation}
indicates that at each step of the renormalization procedure energy and momentum are related by the 
Bogoliubov dispersion relation. 

Contrary to the corresponding renormalization group equations of the ideal
 gas (compare with Eqs.(\ref{idealRGb}))
Eqs.(\ref{RG}) involve two nonlinear coupled equations of motion 
for the scaled parameters $M(l)$ and $G(l)$.
%
The corresponding mean-field results would be obtained by
setting the fluctuations around the mean-field equal to zero. In this case,
both $A[M(l),b(l)]$ and $B[M(l),b(l)]$ vanish and the equations of motion
for $M(l)$ and $G(l)$ decouple.
Furthermore, comparison with Eq.(\ref{idealRGb})
reveals that in the mean-field approximation
the equation for the scaled chemical
potential $M(l)$ is identical to the corresponding equation for the ideal gas which just describes the trivial scaling of $M(l)$.
It should be mentioned that, contrary to the trivial scaling of $M(l)$, 
the trivial scaling of the dimensionless coupling parameter $G(l)$ 
depends on the dimensionality of the problem. 

The quantity $d(l)$ appearing in
these RG equations describes effects of confinement as long as 
they can be modelled by periodic boundary conditions.
These periodic boundary conditions lead to a quantization of the momentum components in the confined directions. 
Thus the free-space integrals over momentum which appear in the continuum limit are replaced by sums of the form
\begin{equation}
\sum_{{\bf m} \in \delta V_{\bf k}}^{'} = V(\Lambda~e^{-l})^{D}
 d(l) dl + O[(dl)^{2}],
\end{equation}
where $d(l)$ characterizes the number of momentum states which are contained
 in an infinitesimal
momentum shell whose thickness is proportional to $dl$ (compare with Fig.\ref{Fig2}).
For example, in the case of confinement of one spatial direction,
say the z-direction, we obtain in the case of a $D$-dimensional problem the expression
\begin{equation}
d(l)= \frac{\Omega_D}{(2\pi)^D}
\frac{\pi + 2\pi[L_z e^{-l} \Lambda/(2\pi)]}{L_z e^{-l}\Lambda}
\label{confi}
\end{equation}
where $L_{z}$ denotes the length of confinement. ($[x]$ denotes the largest integer which is less or
equal to x.)
Taking the limit $L_{z} \to \infty$, Eq.(\ref{confi}) yields $d(l)=\Omega_D/(2 \pi)^{D}$ and we obtain again
the free-space 
renormalization group equations. For the special case of three spatial dimensions these free-space results
reduce to those of Bijlsma and Stoof \cite{StoofBijlsmaRen}.

\subsection{Universal Critical Properties}

For an investigation of the universal critical properties 
of the interacting Bose gas
we have to determine the unstable fixed point of the renormalization
 group equations Eqs.(\ref{RG}) and its corresponding linearized 
flow equations.
For this purpose we have to study Eqs.(\ref{RG}) 
in the limit $l\to\infty$ in which the
effective temperatures are large, i.e. $\beta(l) = \beta_0 ~e^{-2l}\ll 1$ 
 \cite{StoofBijlsmaRen}.
In this limit the quantities $A[M(l),b(l)]$ and $B[M(l),b(l)]$ 
simplify to the expressions
\begin{eqnarray}
A[M(l),b(l)]~&\to&
-2\frac{\epsilon_>^3 + 5 M(l)\epsilon_{>}^{2} + 2M^2(l)\epsilon_{>} +2 M^3(l) \epsilon_{>}}{b(l) \epsilon_>^2[\epsilon_> + 2 M(l)]^2},\nonumber\\
B[M(l),b(l)]~
&\to&
-\frac{5\epsilon_>^2 + 2\epsilon_> M(l) + 2M^2(l)}{b(l) \epsilon_>^2 [\epsilon_> + 2M(l)]^2}.
\end{eqnarray}
Therefore, the equation of motion for $M(l)$ does not couple 
directly to $G(l)$ but to the
modified, scaled coupling $\overline{G}(l) \equiv d(l)[G(l)/b(l)]$. 
In the absence of any confinement, i.e.
for $d(l) = \Omega_D/(2\pi)^D$, the trivial scaling of this
scaled coupling is given by $\overline{G}(l) = \overline{G}(0)~e^{-(D-4)l}$ so that this quantity is relevant for dimensions $D < 4$
contrary to the variable $G(l)$ itself whose trivial scaling indicates that it is relevant for dimension $D < 2$.
However, in the case of confinement this trivial scaling changes.
If one degree of freedom is confined, for example, Eq.(\ref{confi}) implies that for sufficiently large
values of $l$ the trivial scaling of the scaled coupling is given by 
$\overline{G}(l) = \overline{G}(0)~e^{-(D - 4 -1)l}$
which typically has a significant influence on the position of the unstable
 fixed point of the renormalization group equations.

\begin{figure}
\begin{center}
\includegraphics[scale=0.4]{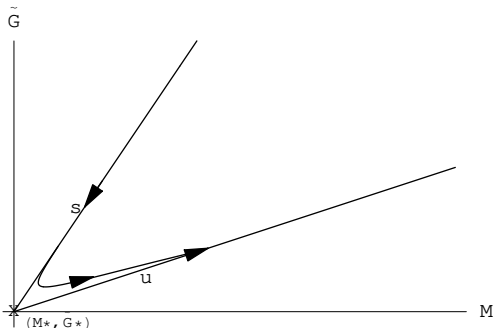}
\caption{Flow of the scaled thermodynamic parameters
 $(M(l),\tilde{G}(l))$ originating in the RG equations Eqs.(\ref{RG})
for $D = 3$. The unstable, fixed point is indicated by a cross and 
the corresponding
stable (s) and unstable (u) manifolds are 
indicated by arrows.
\label{Fig3}}
\end{center}
\end{figure}

Let us consider the physically important case of $D=3$ 
in the absence of confinement in more detail. 
It is straightforward to determine the unstable fixed point 
for the three dimensional case, i.e. 
\begin{eqnarray}
&& M_{*}=1/2 \nonumber \\
&& \tilde{G}_{*}\equiv(G/b)_{*}=\pi^{2}/2
\label{fixedpoint}
\end{eqnarray}
with $\tilde{G}(l) = G(l)/b(l)$.
 The parametric plot of a typical trajectory $\tilde{G}(l)$ versus $M(l)$ is 
drawn in Fig.\ref{Fig3} together with the fixed point and the associated 
stable (s) and unstable (u) manifolds. The eigenvectors corresponding to 
the stable and unstable manifolds are obtained by linearizing the 
RG equations Eqs.(\ref{RG}) around the fixed point. 

\begin{eqnarray}
\frac{d \Delta M(l)}{dl} &=&2 \Delta M(l) -
\frac{2}{\pi^2}\Delta \tilde{G}(l) \nonumber\\
\frac{d \Delta \tilde{G}(l)}{dl} &=& -\Delta \tilde{G}(l) +
\frac{2\pi^2}{3} \Delta M(l) \nonumber
\label{linearized}
\end{eqnarray}
with
$\Delta M=M(l)- M_{*}$ and
$\Delta \tilde{G}=\tilde{G}(l)- \tilde{G}_{*}$.
The eigenvalues and eigenvectors of this set of linear equations are given by
\begin{equation}
\lambda_{\pm} = (3 \pm \sqrt{33})/6
\label{eigenvalues}
\end{equation}
and 
\begin{equation}
(\Delta M, \Delta \tilde{G})_{\pm} = (\frac{9\pm \sqrt{33}}{4\pi^2}, 1).
\end{equation}
According to Eq.(\ref{corrlength})
the positive eigenvalue $\lambda_+ = (3+\sqrt{33})/6$
describes the rate of increase of the relevant variable and determines
the critical exponent for the correlation length, i.e.
\begin{equation}
\nu\equiv 1/\lambda_{+}= \frac{6}{3+\sqrt{33}}\approx 0.686.
\label{critexp}
\end{equation}
as already shown in \cite{StoofBijlsmaRen}. 
This critical exponent compares well with the known result of 
$\nu = 0.67$ \cite{Zinn}.
The corresponding mean-field value for the critical exponent is
$\nu_{{\rm MF}}= 1/2$, exactly the same as for the ideal gas.
 From these considerations it is apparent that 
taking into account fluctuations beyond the mean-field approach is crucial
for describing the critical behaviour of the system.

\subsection{Non-Universal Critical Properties}

The renormalization group equations can also be used to calculate
non-universal properties
of the interacting Bose gas at the critical temperature, for example.
For this purpose one has to solve the 
renormalization group equations of Eq.(\ref{RG})
along the stable manifold of the unstable, fixed point $(M_*,\tilde{G}_*)$.
As an example 
let us consider the critical behaviour
of the (spatially averaged) second order coherence factor $g^{(2)}$ in the
physically interesting case of $D=3$ \cite{Alber}. 
This second order coherence factor is well-known in quantum optics from the correlation experiments of Hanburry-Brown and Twiss
\cite{Brown,MilonniEberly} on electromagnetic radiation from distant stars.
It describes the (spatially averaged) bunching properties of bosons, i.e. the tendency of bosons to
be found at the same position in space, and is defined as follows
\begin{eqnarray}
{\rm g}^{(2)}(0) &\equiv& \frac{\frac{1}{V}\int_V d^3 {\bf x} 
\langle \hat{\psi}^{\dagger}({\bf x}) \hat{\psi}^{\dagger}({\bf x})
\hat{\psi}({\bf x}) \hat{\psi}({\bf x})\rangle}
{[\frac{1}{V}\int_V d^3 {\bf x} \langle \hat{\psi}^{\dagger}({\bf x})
 \hat{\psi}({\bf x})\rangle ]^2}.
\label{g2}
\end{eqnarray}
This quantity can be evaluated from the thermodynamic function $F(M_0,b_0,V)$
by appropriate derivatives with respect to the chemical potential or with respect to the coupling strength $g$ \cite{Alber}.

\begin{figure}
\begin{center}
\includegraphics[angle=270,scale=0.4]{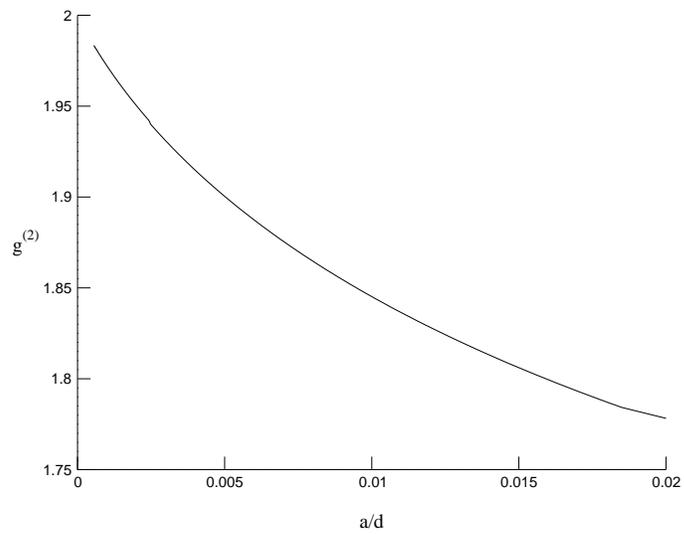}
\caption{Dependence of the second order coherence factor $g^{(2)}$ on the scattering length $a$ at the critical
temperature. The mean distance between the bosons is denoted $d \equiv (V/N)^{1/3}$. \label{Fig4}}
\end{center}
\end{figure}

At the critical temperature
the dependence of this second order coherence factor on the 
scattering length $a$
is depicted in Fig.\ref{Fig4}.
It is apparent that with increasing scattering length 
${\rm g}^{(2)}$ decreases. This is consistent with the intuition 
that with increasing
repulsive interaction bosons tend to avoid each other so that 
the probability to be
found at the same position in space decreases.
In the non-interacting case, i.e. for $a = 0$ and $\mu = 0$, 
this second order coherence factor assumes the 
value of $2$ consistent with the well-known behaviour of a 
photon gas or chaotic field.

\begin{figure}
\begin{center}
\includegraphics[angle=270,scale=0.4]{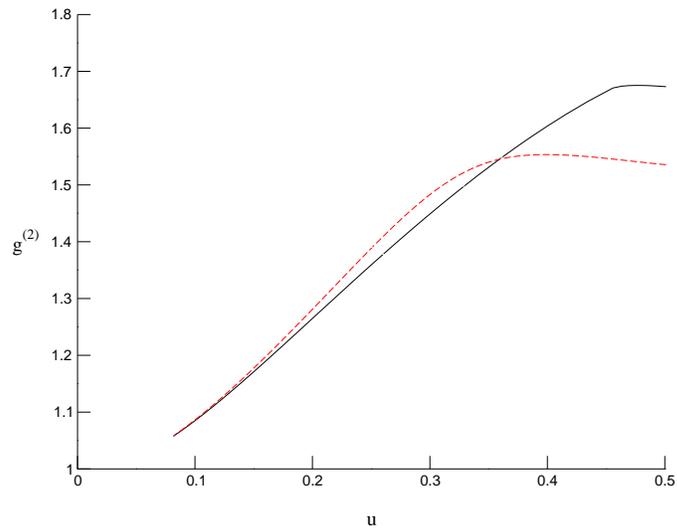}
\caption{ Isothermal dependence of the second order coherence factor $g^{(2)}$ on the scaled
volume $v = (V/N)\lambda^{-3}$ for a thermal de Broglie wave length $\lambda_{\rm th} = 25 a$.
The continuous line corresponds to large confinement length $L_{z} 
/\lambda_{{\rm th}} \gg 1$ and the dashed line to a small confinement length 
 $L_{z}  = 1.2 \lambda_{{\rm th}}$. For a large confinement length there is
 clear evidence of a second order phase transition at  $ v \equiv (V/N)\lambda_{{\rm th}}^{-3} = 0.456$
 which is smoothed out with decreasing confinement length.
 \label{Fig5}}
\end{center}
\end{figure}

Finally, let us consider the influence of confinement in one spatial direction on the second order coherence factor
${\rm g}^{(2)}$. 
For this purpose we insert expression (\ref{confi}) for the density of states into the renormalization
group equations (\ref{RG}).
The resulting dependence of ${\rm g}^{(2)}$ on mean distance between
 the interacting bosons at fixed temperature
is depicted in Fig.\ref{Fig5}.
Temperature and scattering length are chosen in such a way that 
$\lambda_{\rm th} = 25 a$ where 
$\lambda_{\rm th} = \sqrt{2\pi\hbar^2/(mk_B T)}$ denotes
 the thermal de Broglie wave length.
In the case of $^{87}{\rm Rb}$ atoms, for example, 
with a scattering length of $a = 5.3 {\rm nm}$ this condition
corresponds to a temperature of $T = 1.98 \mu K$.
The full curve shows a case in which the characteristic 
length of confinement $L_z$ is much larger than
the thermal de-Broglie wave length. 
In this case the characteristic signature of a second order
phase transition is apparent at a critical scaled volume of
 magnitude $v \equiv (V/N)\lambda_{\rm th}^{-3} = 0.456$.
As soon as the characteristic length of confinement $L_z$
 becomes comparable to the thermal de Broglie wave length
this pronounced signature of a second order phase transition disappears.

\section{Summary}
The renormalization group constitutes a powerful method for evaluating 
thermodynamic properties
of interacting quantum gases beyond the limitations of mean-field theory.  
The previous considerations demonstrate that effects of confinement can be
 described by already existing momentum-space renormalization group 
techniques as long as
confinement can be modelled by periodic boundary conditions. 
Within such an approach confinement leads
to a modification of the density of the eliminated states in the Kadanoff
 transformation, thus affecting
the flow equations of the renormalization group. Typically, 
these modifications change the fixed points of the renormalization group
 equations and thus influence the 
critical properties in a significant way. The presented treatment 
of effects of confinement by periodic boundary conditions constitutes a first
step towards the more complicated final goal of obtaining 
an understanding of thermodynamic properties of trapped interacting
 quantum gases in realistic, smooth particle traps.

\section*{Acknowledgement}
This work is supported by the DFG within the Forschergruppe 'Quantengase'. The authors wish to thank I.~J.~R. Aitchison, M. Cirone, M. Kalinski, V. Kozlov
 and W.~P. Schleich for many valuable discussions. 

\newpage


\newpage

\end{document}